\documentclass[pra,twocolumn,amsmath,amssymb]{revtex4}
\usepackage{amsfonts}
\usepackage{latexsym}
\usepackage{graphicx}
\begin{document}
\title{Stabilization of the number of Bose-Einstein condensed atoms 
in evaporative cooling via three-body recombination loss}
\author{Makoto Yamashita and Tetsuya Mukai} 
\affiliation{NTT Basic Research Laboratories, NTT Corporation, 
3-1, Morinosato-Wakamiya, Atsugi-shi, Kanagawa 243-0198, Japan 
}
\begin{abstract}
The dynamics of evaporative cooling of magnetically 
trapped $^{87}$Rb atoms is studied on the basis of the quantum 
kinetic theory of a Bose gas. We carried out the quantitative 
calculations of the time evolution of conventional evaporative 
cooling where the 
frequency of the radio-frequency magnetic field is swept exponentially. 
This ^^ ^^ exponential-sweep cooling" is known to become inefficient 
at the final stage of the cooling process due to a serious three-body 
recombination loss. We precisely examine how the growth of a 
Bose-Einstein condensate depends on the experimental parameters 
of evaporative cooling, such as the initial number of trapped atoms, 
the initial temperature, and the bias field of a magnetic trap. 
It is shown that three-body recombination drastically depletes 
the trapped $^{87}$Rb atoms as the system approaches the quantum 
degenerate region and the number of   condensed atoms 
finally becomes insensitive to these experimental parameters. 
This result indicates that the final number of condensed 
atoms is well stabilized by a large nonlinear three-body 
loss against the fluctuations of experimental conditions 
in evaporative cooling. 
\end{abstract}
\maketitle
\par \bigskip
\section{Introduction}
Evaporative cooling is the key experimental technique in 
recent successful demonstrations of Bose-Einstein condensation (BEC) 
with trapped neutral atoms \cite{Rb,Li,Na,H,He}. 
The cooling mechanism is understood to be 
that evaporation caused by atomic 
collisions selectively removes the energetic atoms and collisional 
rethermalization simultaneously lowers the temperature of atoms 
remaining in a trap \cite{eva,wal}. 
The utilization of interatomic elastic collisions makes evaporative 
cooling highly efficient. 
Ultra-low temperatures of the order of 
submicro-Kelvin have been achieved in BEC experiments 
by applying evaporative cooling to laser-cooled atoms. 
However, in the case of real atoms, 
other undesirable collisions occur, such as elastic collisions 
with background gas and inelastic collisions due to 
dipolar relaxation and three-body recombination \cite{eva,wal}. 
These collisions result in the loss of trapped atoms 
and seriously reduce the effectiveness of evaporative cooling. 
The efficiency of evaporative cooling is therefore 
determined by the competition between 
evaporation and the loss due to the undesirable collisions. 
\par
Several theoretical analyses based 
on a classical gas model 
have addressed the complicated dynamics of evaporative cooling 
\cite{eva,wal,Lovel,Tommila,Hess,Doyle,Davis,Luit,Surkov,Sack,
Kris,1D_ev}.   
The results have been applied to the optimization 
of experimental conditions \cite{Sack,1D_ev} and have provided 
useful guidelines for reaching the BEC transition point experimentally. 
However, the classical theories fail in the quantum degenerate 
region at low temperatures. 
Bose statistics of atoms strongly modifies the dynamics of 
evaporative cooling there: 
interatomic elastic collisions are enhanced by 
the stimulated scattering process, 
and the event rates of inelastic collisions due to dipolar 
relaxation and three-body recombination  
are reduced by the coherence of identical condensed atoms 
\cite{kagan,Rb_3body_F=1,Rb_3body_F=2}. 
Recent theoretical works  
\cite{Holland,Wu,Jaksch,Yama_1,Holland_2,Yama_H} 
have proved that the quantum kinetic theory taking into account 
the bosonic nature of atoms is a powerful 
tool to gain a comprehensive understanding of evaporative 
cooling in BEC experiments. 
\par
In our previous paper \cite{Yama_opti}, we studied the optimization 
of evaporative cooling in a magnetically trapped $^{87}$Rb system on 
the basis of the quantum kinetic theory. 
We demonstrated that the acceleration of evaporative 
cooling around the BEC transition point is very effective against the 
loss of trapped atoms due to three-body recombination. 
The number of condensed atoms is largely 
enhanced by simply optimizing a sweeping function 
of the frequency of radio-frequency (rf) magnetic field. 
On the other hand, this optimized cooling is determined  
for a fixed certain experimental condition and its cooling trajectory is 
sensitive to the experimental parameters of evaporative cooling,  
such as the initial number of trapped atoms and the initial temperature.   
The optimized evaporative cooling therefore might show the 
fluctuations of the final number of condensed atoms due to some 
variations of these experimental parameters inherent in the real 
experiments. 
In current BEC experiments, the number of condensed atoms 
is usually measured by the destructive time-of-flight method. 
The stable production of condensates with small 
number-fluctuations is thought to be another essential property of 
evaporative cooling. 
\par
In this paper, we quantitatively analyze the dynamics of the conventional 
evaporative cooling where rf-frequency is swept exponentially. 
This ^^ ^^ exponential-sweep cooling" is known to become inefficient 
at the last stage of the cooling process due to three-body recombination 
loss. 
We precisely examine the dependence of condensate 
growth on the experimental parameters, such as 
the initial number of trapped atoms, the initial temperature, and 
the bias field of magnetic trap. 
We demonstrate that three-body loss drastically depletes 
the trapped $^{87}$Rb atoms around the BEC transition point 
and that the number of condensed atoms finally becomes 
insensitive to these experimental parameters. The final number of 
condensed atoms is well stabilized by a large nonlinear three-body loss 
against the fluctuations of experimental conditions. 
\par\bigskip

\section{Quantum kinetic theory of evaporative cooling}
In this section, we briefly mention our theoretical framework 
(See Ref.\ \cite{Yama_1}). 
During evaporative cooling, the magnetic trapping potential $U({\bf r})$ is 
truncated at the energy, which is determined by the frequency of the 
applied rf-magnetic field. 
The thermalized distribution of the noncondensed atoms in such a 
truncated potential is well described by the truncated Bose-Einstein 
distribution function such that 
\begin{equation}
\tilde f({\bf r},p)=\frac{1}{\exp(\epsilon_p/k_BT)/\tilde \xi({\bf r})-1}
\ \Theta (\tilde A({\bf r})-\epsilon_p), 
\label{truncated}
\end{equation}
where $\epsilon_p=p^2/2m$ is the kinetic energy of atoms with 
momentum $p$ and mass $m$, $T$ is the temperature, and 
$\Theta (x) $ is the step function. We add tilde to the notations 
of the quantities obtained through this truncated distribution function. 
The function $\tilde \xi ({\bf r})$ represents the local fugacity and 
includes the mean-field potential energy such that  
$\xi ({\bf r})=\exp\{[\mu-U({\bf r})-2v\tilde n({\bf r})]/k_BT\}$, where 
$\mu $ is the chemical potential, $v=4\pi a \hbar^2$ the 
interaction strength of atoms in proportional to the $s$-wave scattering 
length $a$, and $\tilde n({\bf r})$ the number density of atoms. 
The step function in Eq.\ (\ref{truncated}) eliminates the momentum state 
whose kinetic energy exceeds the effective potential depth 
$\tilde A({\bf r})=\epsilon_t-U({\bf r})-2v\tilde n({\bf r})$, 
where $\epsilon_t$ is the truncation energy of a magnetic trapping potential. 
The number density of atoms 
$\tilde n({\bf r})$ and 
the internal energy density $\tilde e({\bf r})$ are evaluated 
through the truncated distribution function in a self-consistent way:
\begin{eqnarray}
\tilde n({\bf r}) & = & \frac{4\pi}{h^3} \int \tilde f ({\bf r}, p) 
\ p^2 dp, \\
\label{den_n}
\nonumber \tilde e({\bf r}) & = & \frac{4\pi}{h^3} \int \epsilon_p \ \tilde f ({\bf r}, p) 
\ p^2 dp +v [\tilde n({\bf r})]^2 \\
&&
+ U({\bf r}) \tilde n({\bf r}). 
\label{den_e}
\end{eqnarray}
Thus both $\tilde n({\bf r})$ and $\tilde e({\bf r})$ are the 
complicated functions of  $T$, $\mu$,  $\epsilon_t$, $U({\bf r})$, and $a$. 
The spatial integrations of these density functions give us the total 
number of atoms, 
$\tilde N=\int\tilde n({\bf r}) d{\bf r}$, and the total internal energy, 
$\tilde E=\int\tilde e({\bf r}) d{\bf r}$.  
After the BEC transition, the number density in the condensed region 
is given by the sum of a condensate $n_0({\bf r})$ and 
saturated noncondensed atoms $\tilde n_n$ such that 
$\tilde n({\bf r})=n_0({\bf r})+\tilde n_n$. 
The condensate $n_0({\bf r})$ is approximated by the Thomas-Fermi 
distribution and $\tilde n_n$ is evaluated from the truncated 
Bose-Einstein distribution function on the condition that 
the local fugacity becomes unity, i.e., $\tilde \xi({\bf r})=1$. 
\par 
During the cooling process, trapped atoms are removed from the 
potential by both evaporation and undesirable collisions. 
The change rates of the total number of atoms, 
$\tilde N$, and the total 
internal energy, $\tilde E$, are thus calculated as the 
sum of the contributions of all processes \cite{Yama_opti}: 
\begin{eqnarray}
\nonumber \frac{d \tilde N}{dt} &=&-\int\ \dot n_{\rm ev}({\bf r})
\ {\rm d}{\bf r}-
\sum_{s=1}^3 \ G_s \int K_s({\bf r}) [\tilde n({\bf r}) ]^s\ {\rm d}{\bf r}  
\\
&&
+\left( \frac{\partial \tilde N }{\partial \epsilon_t}\right)_{T, \mu} 
\dot \epsilon_t, 
\label{difeq_N}
\\
\nonumber  \frac{d \tilde E}{dt} &=&-\int\ \dot e_{\rm ev} ({\bf r})\ {\rm d}{\bf r} -
\sum_{s=1}^3 
G_s \int K_s({\bf r}) \tilde e({\bf r}) [\tilde n({\bf r})]^{s-1}\ {\rm d}{\bf r}
\\ 
&&
+\left( \frac{\partial \tilde E }{\partial \epsilon_t}\right)_{T, \mu} \dot 
\epsilon_t.
\label{difeq_E}
\end{eqnarray}  
The $\dot n_{\rm ev}({\bf r})$ and $\dot e_{\rm ev}({\bf r})$ denote 
the evaporation 
rates of density functions derived from a general collision integral of 
a Bose gas system.  
In Eqs.\ (\ref{difeq_N}) and (\ref{difeq_E}), 
we adopt the opposite sign of the notations of these 
evaporation rates, which were 
defined in our previous paper \cite{Yama_1}. 
Parameters $G_1$,  $G_2$, and $G_3$ are the decay rate constants of 
trapped atoms due to the background gas collisions, dipolar relaxation, 
and three-body recombination, respectively. 
The function $K_s({\bf r})$ represents the correlation function, which 
describes the $s$-th 
order coherence of trapped atoms, and we use the 
expressions of $K_s$ for an ideal Bose 
gas system \cite{kagan,Rb_3body_F=1,Rb_3body_F=2}. 
The terms proportional to the change rate of truncation energy, 
$\dot \epsilon_t=d\epsilon_t/dt$, give the contribution of extra atoms 
that ^^ ^^ spill over" when $\epsilon_t$ decreases continuously in forced 
evaporative cooling \cite{Kris,Yama_opti}.  
\par
The time-evolution was calculated numerically 
by assuming that the system always stays in quasi-thermal equilibrium 
state and is described by the truncated Bose-Einstein distribution 
function. 
This quick rethermalization approximation is appropriate for the slow 
evaporative cooling normally adopted in BEC experiments 
\cite{Yama_1,Luit}. 
\par \bigskip
\section{Numerical results and discussins}
In this section, we show the numerical results of 
the time-evolution calculations of evaporative cooling corresponding to 
the several cases of experimental conditions. 
To carry out the quantitative calculations, we employed the 
relevant parameters 
corresponding to the experimental setup at Gakushuin University 
\cite{Gakushuin}. 
The cloverleaf magnetic trap characterized by bias field $B_0$, 
radial gradient $B^{'}$, 
and axial curvature  $B^{''}$ was modeled by
a magnetic field of the Ioffe-Pritchard type such that 
$B(r,z)=\sqrt{(\alpha r)^2+(\beta z^2+B_0)^2}$,  
where $\alpha=\sqrt{B^{'2}-B^{''}B_0/2}$ and $\beta=B^{''}/2$. 
We used the typical values adopted experimentally, such as 
$B_0=1.6\ {\rm G}$, $B^{'}=169\ {\rm G/cm}$,  and $B^{''}=148
\ {\rm G/cm^2}$. 
The corresponding trapping frequencies are $\omega_r=2 \pi 
\times 170$\ Hz in the radial direction 
and $\omega_z=2 \pi \times 15.5$\ Hz in the axial direction. 
\par
The frequency of the rf-magnetic field, $\nu_{\rm rf}$, is assumed 
to be swept exponentially 
from 35 to 1.13 MHz in 30 s such that $\nu_{\rm rf}(t)=\nu_a
+(\nu_b-\nu_a) \ e^{-t/\tau}$ 
with $\nu_a=1.13$ MHz, $\nu_b=35$ MHz, and $\tau=3.69$ s. 
This exponential-sweep cooling is 
commonly adopted in BEC experiments \cite{Rb_3body_F=2}, and 
we employed this functional form in the following all numerical calculations. 
Furthermore, we set the collisional parameters of $^{87}$Rb atoms 
in the trapped state of ${\rm F=2, m_F} =+2$ as follows: 
s-wave scattering length $a=5.8$ nm, loss rate 
constants due to 
collisions with background gas $G_1=0.01$ s$^{-1}$, dipolar relaxation 
$G_2=1.0 \times 10^{-15}$ cm$^3$s$^{-1}$ \cite{dipolar}, and 
three-body relaxation 
$G_3=1.1\times 10^{-28}$ cm$^6$s$^{-1}$ \cite{Rb_3body_F=2}. 
\par
\subsection{Dependence on the initial number of trapped atoms}
Considering that the initial temperature is strongly restricted by 
the limit of the laser cooling done before evaporative cooling 
\cite{laser_cool}, we first carried out the time-evolution 
calculations for several initial numbers of atoms 
with the initial temperature fixed. 
Under the typical initial temperature of $T_i=380 \ \mu$K \cite{Gakushuin},  
we varied the initial total number of atoms from $N_i=1.5 \times 10^8$ 
to $3 \times 10^9$ and investigated how 
the final number of condensed atoms depends on these values. 
\par
%
\begin{figure}[t]
\centerline{\includegraphics[height=8.5cm]{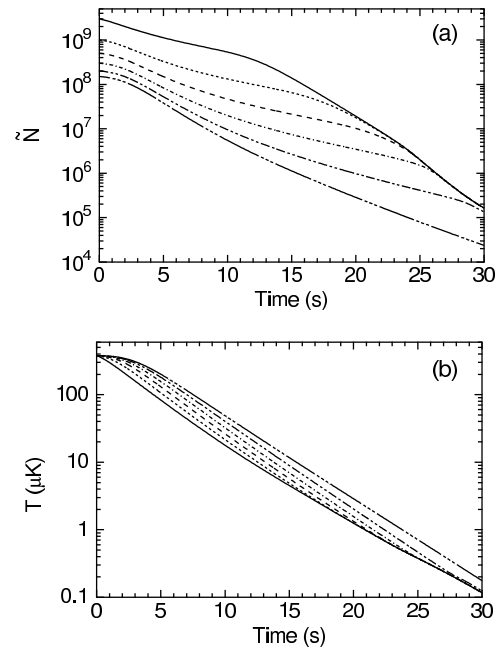}}
\caption{ \label{NT_Ni}Time evolution of (a) total 
number of atoms $\tilde N$ and 
(b) temperature $T$ for several initial conditions of evaporative cooling: 
the initial number of trapped atoms $N_i$ is varied 
from $1.5 \times 10^8$ to $3 \times 10^9$ 
under  the fixed initial temperature of $T_i=380 \ \mu$K. 
In both figures, the curves correspond to cases with 
$N_i=1.5 \times 10^8$ 
({\unitlength 1pt \thicklines
\put(0,3){\line(1,0){8}} \multiput(10,3)(4,0){3}{\line(1,0){2}}
\put(22,3){\line(1,0){8}}}\hspace{30pt}),
$2 \times 10^8$ 
({\unitlength 1pt \thicklines \put(0,3){\line(1,0){7}}
\multiput(9,3)(4,0){2}{\line(1,0){2}} \put(17,3){\line(1,0){7}}
}\hspace{24pt}),
$3 \times 10^8$ 
({\unitlength 1pt \thicklines \put(0,3){\line(1,0){8}}
\multiput(10,3)(4,0){1}{\line(1,0){2}}
\put(14,3){\line(1,0){8}}
}\hspace{22pt}),
$5 \times 10^8$ 
({\unitlength 1pt \thicklines
\multiput(0,3)(7,0){3}{\line(1,0){3.5}}
}\hspace{20pt}),
$1 \times 10^9$ 
({\unitlength 1pt \thicklines
\multiput(0,3)(2,0){10}{\line(1,0){1}}
}\hspace{20pt}), and 
$3 \times 10^9$ 
({\unitlength 1pt \thicklines
\put(0,3){\line(1,0){20}}
}\hspace{20pt}).}
\end{figure}%
Figures\ \ref{NT_Ni}(a) and 1(b) respectively show the time-evolution 
of the total number of atoms $\tilde N$ and temperature $T$ for 
several  initial conditions of evaporative cooling. 
We can see that both $\tilde N$ and $T$ decrease almost 
exponentially over time. Furthermore, in Fig.\ \ref{NT_Ni}(a), 
four of the curves from the solid line to the dash-dotted line 
corresponding to $N_i=3 \times 10^8 \sim 3 \times 10^9$ 
merge into one, 
which means ten times more initial atoms are completely lost 
during the 30-s cooling.
The time-evolution of temperature in Fig.\ \ref{NT_Ni}(b) also 
shows the same feature. 
The final results of exponential-sweep evaporative cooling become 
insensitive to the initial number of atoms  when cooling starts 
from more than $3 \times 10^8$ atoms.  
\par 
%
\begin{figure}[h]
\centerline{\includegraphics[height=5cm ]{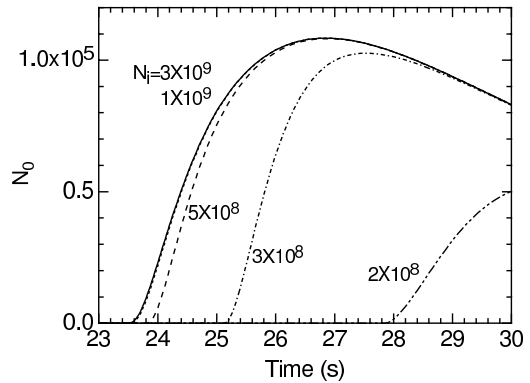}}
\caption{\label{N0_Ni} Time evolution of the number of 
condensed atoms $N_0$ for 
the several  initial numbers of trapped atoms $N_i$. 
We use the same lines as in Fig.\ \ref{NT_Ni} in terms of the 
corresponding $N_i$ values. 
The solid line ($N_i=3 \times 10^9$) and the dotted line 
($N_i=1 \times 10^9$) almost coincide with each other.  
For $N_i=1.5 \times 10^8$, the system can not reach the  
BEC transition point and condensed atoms are not produced. 
}
\end{figure}%
Figure \ref{N0_Ni} shows the dependence of the time evolution of the 
number of condensed atoms, $N_0$, on the initial number of 
trapped atoms $N_i$. 
For $N_i=3 \times 10^8 \sim 3 \times 10^9$,  
we finally obtain the same $N_0$ value of 
$8.3 \times 10^4$ with the same condensate 
fraction of $N_0/\tilde N=51\%$ after 30-s evaporative cooling, 
even though the BEC transition occurs at a different time. 
This implies that, when we apply exponential-sweep evaporative cooling,  
even a one-order change of the initial total number 
of trapped atoms does not affect the final number of condensed atoms. 
Since $N_i =(3 \sim 5) \times 10^8 $ is normally  achieved 
in current BEC experiments with $^{87}$Rb, 
it is concluded that the final number of condensed atoms 
can be well stabilized against the fluctuations of the initial number of atoms 
in the experiments. 
\par
\begin{figure}[h]
\centerline{\includegraphics[height=5cm ]{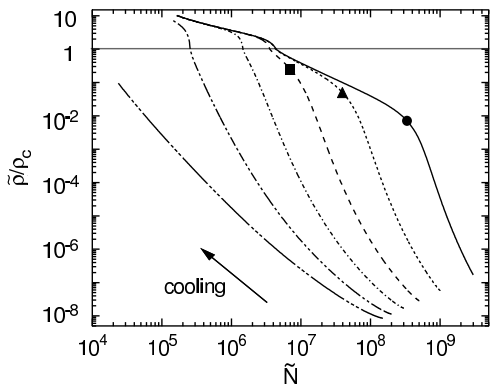}}
\caption{\small \label{trajectory} Cooling trajectories through phase space 
for several initial conditions of evaporative cooling: 
the initial number of trapped atoms $N_i$ is varied under  
the fixed initial temperature of $T_i=380 \ \mu$K. The phase-space 
density $\tilde \rho $ is normalized by the 
critical value $\rho_c=2.612$. The BEC transition occurs when each 
trajectory passes the horizontal line where $\tilde \rho/\rho_c=1$. 
We use the same lines as in Fig.\ \ref{NT_Ni} in terms of the 
corresponding $N_i$ values. 
The arrow shows the direction of time. 
The symbols on the solid line (circle), dotted line (triangle), 
and dashed line (square) indicate the points where the evaporative 
cooling 
becomes significantly inefficient. 
}
\end{figure}%
Here we show the results of the cooling trajectory through phase 
space and provide clear information about the efficiencies of present 
exponential-sweep cooling. 
In a nonuniform gas system, 
the phase-space density is evaluated at the peak position as 
$\tilde \rho=\tilde n({\bf 0})\lambda^3$, 
where $\lambda=h/\sqrt{2\pi m k_B T}$ is the thermal de Broglie 
wavelength with atomic mass $m$. 
In Fig.\ \ref{trajectory}, we plot the cooling trajectories of the 
present evaporative cooling in the $\tilde N$-$\tilde \rho$ plane.  
At the early stage of cooling, all trajectories are well 
separated according to 
the initial conditions and evaporative cooling works efficiently. 
However, as the system approaches the quantum 
degenerate region where the phase-space density exceeds 
the critical value of $\rho_c=2.612$, three of the trajectories 
from the solid line to the dashed line bend towards horizontal 
around the respective turning points indicated by the symbols, 
showing the cooling efficiency drastically decreases there.  
As expected from the results in Fig.\ \ref{NT_Ni}, three of the 
trajectories merge and finally coincide with the dash-dotted line 
corresponding 
to $N_i=3 \times 10^8$. 
On the other hand, when $N_i \leq 2 \times 10^8$, the final results 
after 30-s evaporative cooling strongly depend on the initial 
number of atoms. 
The system can not reach the BEC transition point 
if the cooling is started from $N_i=1.5 \times 10^8$. 
\par
\begin{figure*}[t]
\begin{center}
\includegraphics[width=13cm ]{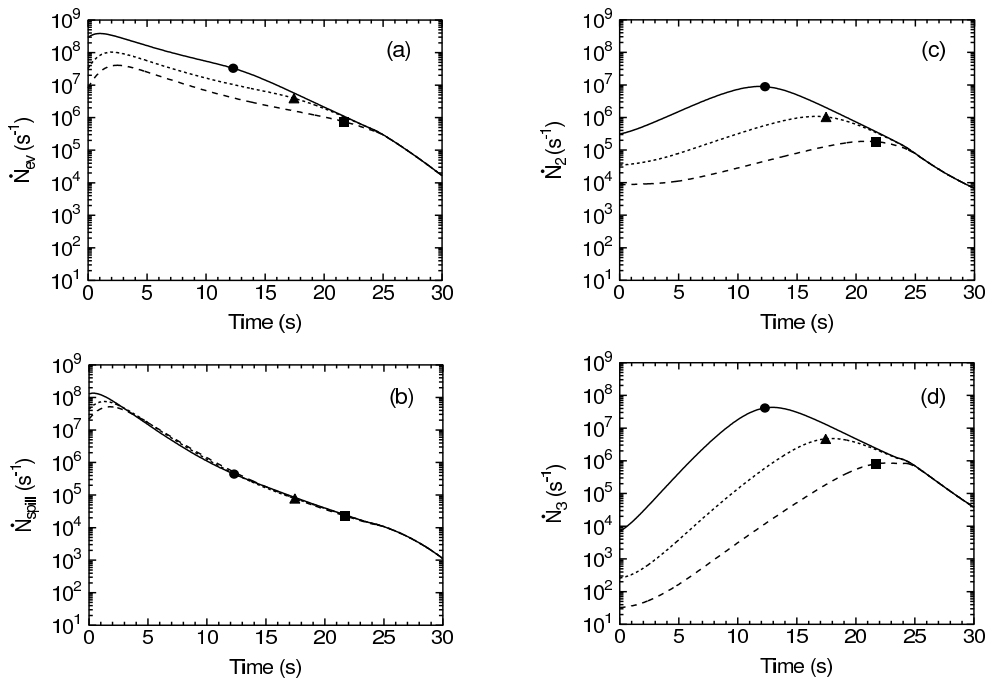}
\caption{\small \label{rate} Time evolution of (a) evaporation 
rate $\dot N_{\rm ev}$, (b) spill rate $\dot N_{\rm spill}$, 
(c)  loss rate due to two-body dipolar relaxation $\dot N_2$, 
and (d) loss rate due to three-body recombination $\dot N_3$. 
We use the same lines as in Fig.\ \ref{NT_Ni} in terms of the 
corresponding $N_i$ values, and show the results for 
the cases of  
$N_i=5 \times 10^8$ 
({\unitlength 1pt \thicklines
\multiput(0,3)(7,0){3}{\line(1,0){3.5}}
}\hspace{20pt}),
$1 \times 10^9$ 
({\unitlength 1pt \thicklines
\multiput(0,3)(2,0){10}{\line(1,0){1}}
}\hspace{20pt}), and 
$3 \times 10^9$ 
({\unitlength 1pt \thicklines
\put(0,3){\line(1,0){20}}
}\hspace{20pt}). 
The symbols in the figures 
correspond to the same points plotted on the cooling 
trajectories in Fig.\ \ref{trajectory}}.
\end{center}
\end{figure*}%
Next, we show the loss rates of trapped atoms during evaporative 
cooling and discuss the origin of the inefficiency of cooling. 
As described in Eq.\ (\ref{difeq_N}), 
the change rates of the total number of atoms decompose into five 
contributions: the evaporation rate $\dot N_{\rm ev}$, 
spill rate $\dot N_{\rm spill}$, loss rates 
due to back-ground gas collision $\dot N_1$, 
two-body dipolar relaxation $\dot N_2$, and 
three-body recombination $\dot N_3$. 
In Fig.\ \ref{rate}, we plot the time evolution of these rates 
for $N_i=3 \times 10^9, 1 \times 10^9$, 
and $5 \times 10^8$. 
The results for $\dot N_1$ are not shown here since 
this rate is given by 
$G_1 \tilde N$ and evaluated straightforwardly from 
Fig.\ \ref{NT_Ni}(a). 
The symbols in the figures correspond to the same points 
plotted on the cooling trajectories in Fig.\ \ref{trajectory} and 
indicate the time where 
the efficiency of evaporative cooling is drastically reduced. 
We see that both $\dot N_{\rm ev}$ and 
$\dot N_{\rm spill}$  \cite{spill}
decrease exponentially over time, 
while $\dot N_2$ and $\dot N_3$  
have peak structures around the symbol on each curve.  
The peak value of $\dot N_3$ is about five times larger 
than that of $\dot N_2$ and 
three-body recombination is the dominant loss mechanism 
for $^{87}$Rb atoms.
By comparing  Fig.\ \ref{rate}(a) with \ref{rate}(d), it is also found 
that the symbol on each curve indicates the time when the 
three-body loss rate exceeds the evaporation rate. 
At the final stage of the present exponential-sweep cooling,  
we understand that the cooling speed becomes so low that 
the three-body loss crucially reduces the cooling efficiency.  
A stabilization of the final number of condensed atoms is 
therefore realized by the combination of slow cooling and large 
three-body loss.
\par
From another viewpoint, the present result also suggests that the 
exponential-sweep of rf-frequency is seriously inefficient when 
we try to make a larger condensate. 
We have thought so far that the improvement of the initial conditions 
results in more condensed atoms even in exponential-sweep 
evaporative cooling.  
However, it is clear in Fig.\ \ref{N0_Ni} that a ten-fold increase of 
the initial number of trapped atoms does not improve the achievable 
number of condensed atoms. 
In order to obtain a larger number of condensed  
atoms, we have to carry out the evaporative cooling quickly at the 
final stage, where the system approaches the quantum degenerate 
region with high density, 
as has been demonstrated quite recently by optimizing the cooling 
trajectories \cite{Yama_opti,Sack,1D_ev}. 
\par
\subsection{Dependence on the initial temperature}
In BEC experiments,  evaporative cooling is normally employed 
after laser cooling \cite{eva,laser_cool}. 
The atoms are transfered from a magneto-optical trap (MOT) 
in laser cooling into a magnetic trap in evaporative cooling.  
The misalignment between these two traps easily causes 
the fluctuations of initial temperature for succeeding evaporative 
cooling process. 
Thus, we next carried out the time-evolution calculations of 
evaporative cooling for several initial temperatures with the initial 
number of trapped atoms fixed. 
We assumed that $5 \times 10^5$ atoms are initially loaded into 
the magnetic trap before evaporative cooling (i.e., $5 \times 10^5$ 
atoms exist in the 
magnetic trap without truncation). 
The initial temperature was then varied from $T_i=280 \ \mu$K to 
$630 \ \mu$K, and we investigated how this variance of initial 
temperature, as well as that of the initial number of atoms, 
is stabilized by a three-body loss. 
\par
\begin{figure}[h]
\centerline{\includegraphics[height=8.5cm]{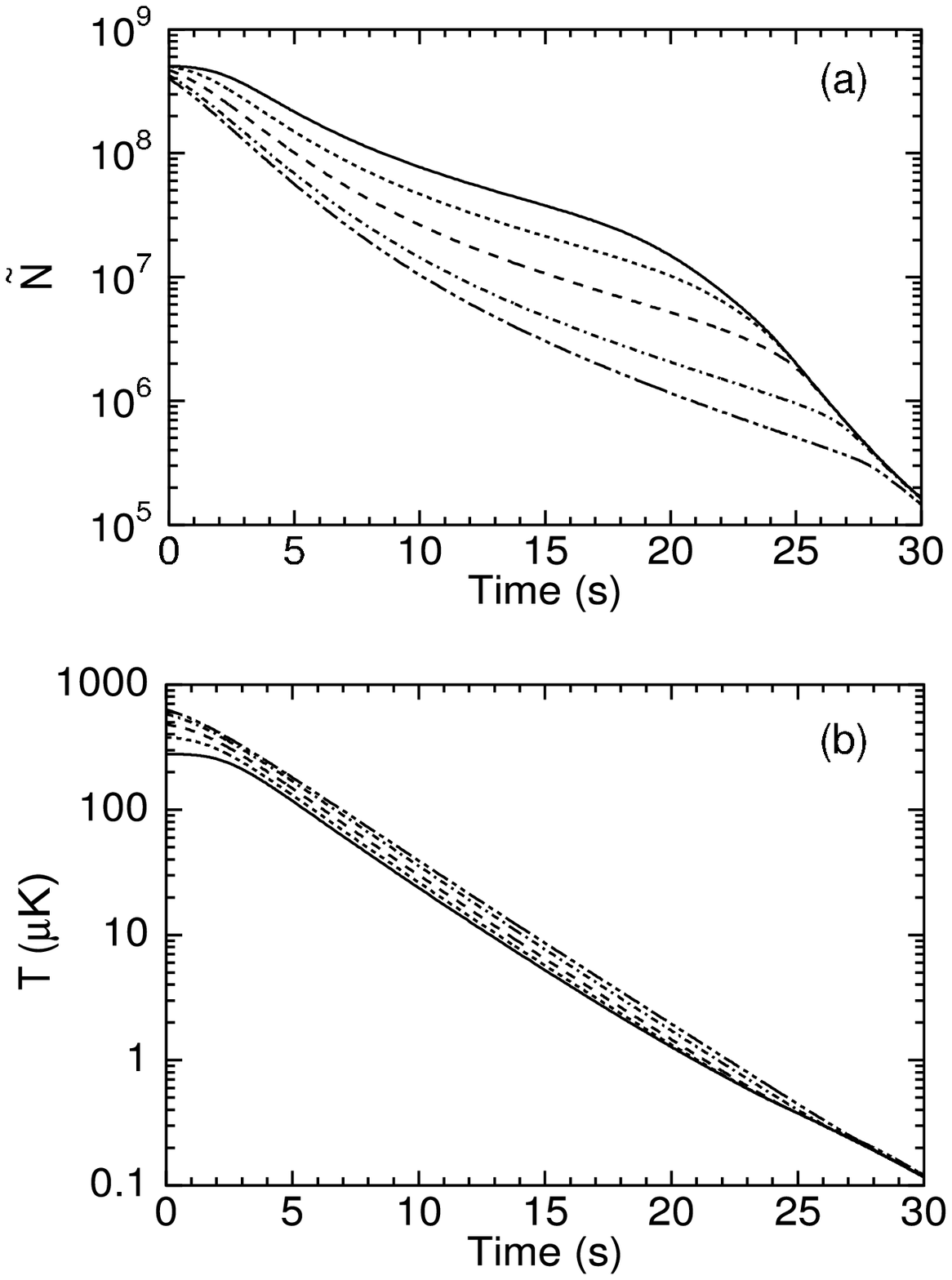}}
\caption{\small \label{NT_Ti}Time evolution of (a) total number of 
atoms $\tilde N$ and (b) temperature $T$ for 
several initial conditions of evaporative cooling: 
the initial temperature $T_i$ is varied under 
a fixed initial number of atoms $N_i\simeq5 \times 10^8$. 
In both figures, the curves correspond to the cases with 
$T_i=630\ \mu$K 
({\unitlength 1pt \thicklines \put(0,3){\line(1,0){7}}
\multiput(9,3)(4,0){2}{\line(1,0){2}} \put(17,3){\line(1,0){7}}
}\hspace{24pt}),
$580\ \mu$K
({\unitlength 1pt \thicklines \put(0,3){\line(1,0){8}}
\multiput(10,3)(4,0){1}{\line(1,0){2}}
\put(14,3){\line(1,0){8}}
}\hspace{22pt}),
$480\ \mu$K
({\unitlength 1pt \thicklines
\multiput(0,3)(7,0){3}{\line(1,0){3.5}}
}\hspace{20pt}),
$380\ \mu$K
({\unitlength 1pt \thicklines
\multiput(0,3)(2,0){10}{\line(1,0){1}}
}\hspace{20pt}), and 
$280\ \mu$K
({\unitlength 1pt \thicklines
\put(0,3){\line(1,0){20}}
}\hspace{20pt}).}
\end{figure}%
Figures \ref{NT_Ti}(a) and \ref{NT_Ti}(b) respectively show the 
time-evolution of the total number of atoms $\tilde N$ and the 
temperature $T$ for several initial conditions of evaporative cooling. 
Both $\tilde N$ and $T$ decrease almost exponentially over time. 
In Fig.\ \ref{NT_Ti}(a), the small difference of $\tilde N$ at $t=0$ 
originates from  the difference of truncation-parameter value 
$\eta=\epsilon_t/(k_BT_i)$ 
for each initial temperature with the same truncation energy 
$\epsilon_t$. We can see that $\tilde N$ curves and $T$ curves for 
the initial temperatures, i.e., from $280 \  \mu$K to $580 \ \mu$K, 
finally merge into one. 
The variance of initial temperatures finally vanishes after 30-s 
exponential evaporative cooling as similarly in Fig.\ \ref{NT_Ni}. 
\par 
\begin{figure}[h]
\centerline{\includegraphics[height=5cm ]{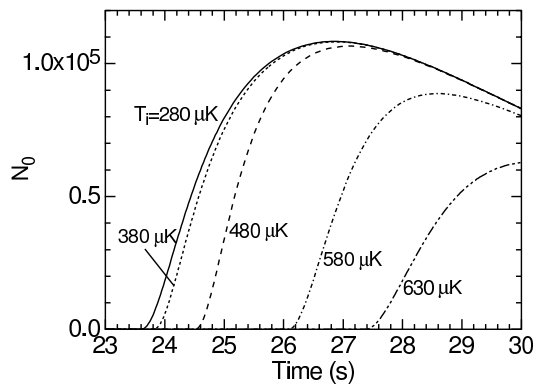}}
\caption{\small \label{N0_Ti} Time evolution of the number of 
condensed atoms $N_0$ for several initial temperarures of 
evaporative cooling. We use the same lines as in Fig.\ \ref{NT_Ti} 
in terms of the corresponding $T_i$ values. 
}
\end{figure}%
In Fig.\ \ref{N0_Ti}, we plot the growth of condensate at the final 
stage of cooling. For $T_i=280 \sim 580 \ \mu$K, the final number 
of condensed atoms at $t=30$ s
takes almost the same value, $N_0=8.3 \times 10^4$, even though 
the BEC transition occurs at different times according to the initial 
temperatures. 
We will obtain the same number of condensate for a wide range of 
initial temperatures. When $T_i$ exceeds $580 \ \mu$K, on the other 
hand, the final value of $N_0$ decreases with increasing $T_i$. 
It is concluded that the number of condensed atoms is 
sufficiently stabilized against a large variance of initial temperatures 
too ($300\ \mu$K in the present calculation). 
We also verified that this stabilization is caused by a combination 
of large three-body loss and slow evaporation at the final stage of 
cooling process, which is exactly the same mechanism as discussed 
in the previous subsection.  
\par 
\subsection{Dependence on the bias field of a magnetic trap}
In BEC experiments, the magnetic trap is usually driven with a
large current of the order of 100 A. 
The current noise causes the fluctuations and drifts of 
the bias field $B_0$ in a magnetic trap, which is the important 
limitation to the stable production of condensates. 
For example, Ref.\ \cite{Rb_3body_F=2} reported that 
the dirift of $B_0$ should be suppressed to less than 3 mG to 
obtain a good accuracy for the number of condensed 
atoms produced experimentally.  
Here we study the dependence of condensate growth on the bias 
field of a magnetic trap. 
The time-evolution calculations of 30-s evaporative cooling were 
carried out for several values of $B_0$ ranging from 1.56 to 
1.614 G, under the initial conditions of $N_i=5 \times 10^8$ and 
$T_i=380\ \mu$K. 
Since we used the same exponential sweeping-function of rf-field 
frequency $\nu_{\rm rf} (t)$, the time dependence of a truncation 
energy, $\epsilon_t(t)$, was equivalent  in these calculations. 
Therefore, at any time during evaporative cooling, a larger bias field 
corresponds to a smaller depth of a truncated trapping potential. 
\par
\begin{figure}[h]
\centerline{\includegraphics[height=5cm ]{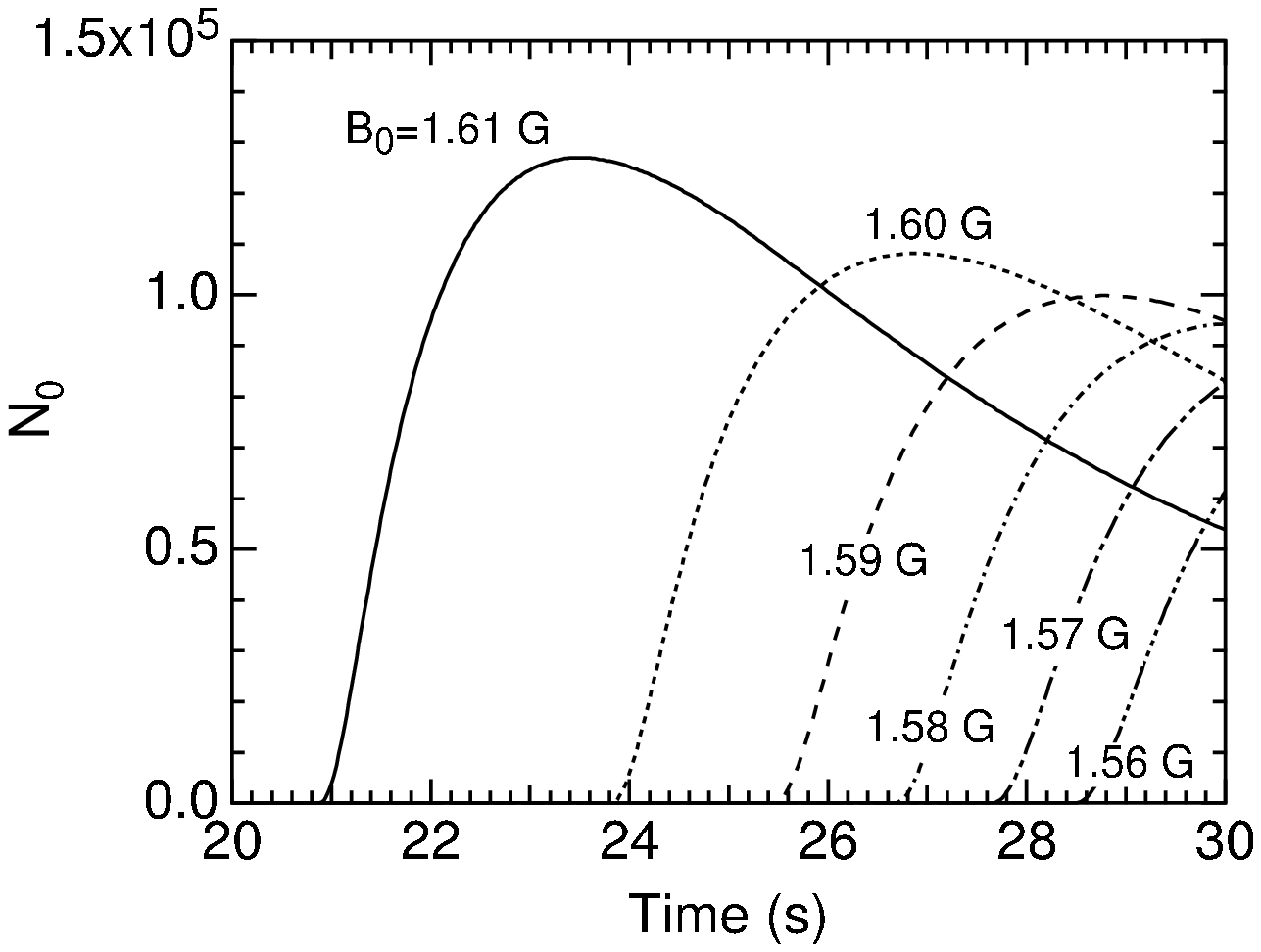}}
\caption{\small  \label{N0_B0} Time evolution of the number of 
condensed atoms $N_0$ for 
several values of bias field.  
The curves correspond to cases with 
$B_0=1.61$\ G
({\unitlength 1pt \thicklines
\put(0,3){\line(1,0){20}}
}\hspace{20pt}) 
$1.60$\ G
({\unitlength 1pt \thicklines
\multiput(0,3)(2,0){10}{\line(1,0){1}}
}\hspace{20pt}), 
$1.59$\ G
({\unitlength 1pt \thicklines
\multiput(0,3)(7,0){3}{\line(1,0){3.5}}
}\hspace{20pt}),
$1.58$\ G
({\unitlength 1pt \thicklines \put(0,3){\line(1,0){8}}
\multiput(10,3)(4,0){1}{\line(1,0){2}}
\put(14,3){\line(1,0){8}}
}\hspace{22pt}),
$1.57$\ G
({\unitlength 1pt \thicklines \put(0,3){\line(1,0){7}}
\multiput(9,3)(4,0){2}{\line(1,0){2}} \put(17,3){\line(1,0){7}}
}\hspace{24pt}),
and 
$1.56$\ G
({\unitlength 1pt \thicklines
\put(0,3){\line(1,0){8}} \multiput(10,3)(4,0){3}{\line(1,0){2}}
\put(22,3){\line(1,0){8}}}\hspace{30pt}).}
\end{figure} 
%
%
\begin{figure}[h]
\centerline{\includegraphics[height=5cm ]{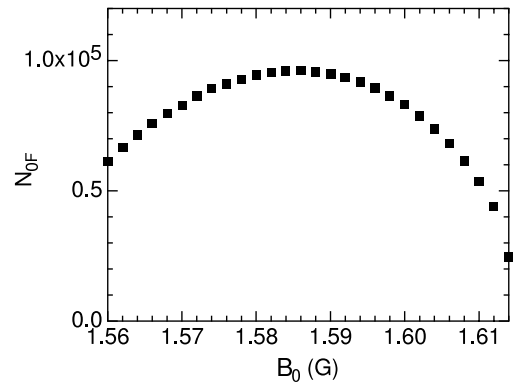}}
\caption{\small \label{B0_dep} Bias field dependence of 
the final number of condensed atoms, $N_{0{\rm F}}$.  
Around the stationary point of $B_0=1.585$\ G, 
we can see that $N_{0{\rm F}}$ is  
sufficiently stabilized against fluctuations of the $B_0$ value 
within about 10 mG . 
}
\end{figure}%
Figure\ \ref{N0_B0} shows the time evolution of the number 
of condensed atoms $N_0$ for the several values of bias field. 
The condensate growth curves strongly depend 
on $B_0$ values as has been expected. 
With larger bias field, the BEC transition occurs 
earlier and the curve shows a peak structure. 
A condensate gradually decays due to three-body loss since 
slow evaporation at the last stage of cooling 
can not preserve the large number of condensed atoms. 
On the other hand, with smaller bias field, 
30-s evaporative cooling is completed before the condensate 
grows sufficiently.
In Fig.\ \ref{B0_dep}, we plot the bias field dependence of
the final number of condensed atoms $N_{0{\rm F}}$. 
It is found that $N_{0{\rm F}}$ is the convex function of $B_0$ 
and the stationary point exists around $B_0 = 1.585$ G. 
When $B_0$ is set at this value, we can expect that the final 
number of condensed atoms is stabilized against 
fluctuations of bias field within about 10 mG.  
Note that this stabilization is also achieved by a large three-body
loss.
\par  
\section{Conclusion}
On the basis of the quantum kinetic theory of a Bose gas, 
we have theoretically studied the dynamics of evaporative cooling 
with magnetically trapped $^{87}$Rb atoms. 
The time-evolutional calculations were carried out quantitatively 
for the conventional exponential-sweep cooling,  
which is known to be inefficient against a three-body recombination 
loss. 
We precisely examined the dependence of a condensate growth 
on the experimental parameters, such as 
the initial number of atoms, the initial temperature, and the bias field 
of a magnetic trap. 
It was shown that three-body recombination drastically depletes the 
trapped atoms as the system approaches the quantum degenerate 
region and the number of condensed atoms finally becomes 
insensitive to these experimental parameters. 
The final number of condensed atoms is well stabilized 
by a large nonlinear three-body loss against fluctuations of 
experimental conditions. 
We can choose the manner of evaporative cooling according to the 
aim of BEC experiments; a conventional exponential-sweep 
cooling if we want a stable production of condensates or an optimized 
one with the rapid sweep of rf-frequency at the final stage of cooling 
procedure if we want to maximize the number of condensed atoms 
\cite {Yama_opti}. 

\begin{acknowledgments}
The authors thank Y. Yoshikawa of the University of 
Tokyo, K. Araki, T. Kuwamoto, T. Hirano of Gakushuin University, 
M. Koashi, N. Imoto of the Graduate University for Advanced Studies, 
M. Mitsunaga of Kumamoto University, 
M. W. Jack of Rice University, T. Hong of the University of Washington, 
and Y. Tokura of 
NTT Basic Research Laboratories for valuable discussions. 
\end{acknowledgments}

\end{document}